# Experimental demonstration of graphene plasmons working close to the near-infrared window

Zhongli Wang,[1,2,§] Tao Li,[1,§] Kristoffer Almdal,[1,2] N. Asger Mortensen,[2,3] Sanshui Xiao,[2,3,*] and Sokol Ndoni[1,2,*]

[1]Department of Micro- and Nanotechnology, Technical University of Denmark, DK-2800 Kgs. Lyngby, Denmark.
[2]Center for Nanostructured Graphene (CNG), Technical University of Denmark, DK-2800 Kgs. Lyngby, Denmark.
[3]Department of Photonics Engineering, Technical University of Denmark, DK-2800 Kgs. Lyngby, Denmark.
[§] Equal contribution
[*]Corresponding author: saxi@fotonik.dtu.dk, sond@nanotech.dtu.dk

**Due to strong mode-confinement, long propagation-distance, and unique tunability, graphene plasmons have been widely explored in the mid-infrared and terahertz windows. However, it remains a big challenge to push graphene plasmons to shorter wavelengths in order to integrate graphene plasmon concepts with existing mature technologies in the near-infrared region. We investigate localized graphene plasmons supported by graphene nanodisks and experimentally demonstrated graphene plasmon working at 2 μm with the aid of a fully scalable block copolymer self-assembly method. Our results show a promising way to promote graphene plasmons for both fundamental studies and potential applications in the near-infrared window.**

Graphene is proved to be a promising material for optoelectronic devices because of its unique electric and photonic properties [1]. Relying on its high carrier mobility and tunability, many fascinating graphene-based optoelectronic devices have been realized, including photodetectors [2], modulators [3, 4], and ultrafast lasers [5]. However, due to its single-atom-layer thickness, graphene interacts with light very inefficiently (only 2.3% absorption) [6], posing challenges and restrictions for graphene-based optoelectronic devices. Many efforts so far have focused on enhancing light-graphene interactions by integrating graphene with dielectric cavity systems and plasmon resonators made from noble metals [7, 8]. An alternative approach for enhancing optical absorption in graphene is to excite graphene-plasmons (GPs) supported by graphene sheets or graphene nanostructures [9, 10]. GPs are confined to volumes much smaller than the diffraction limit, thus facilitating the optical absorption in the single-layer graphene. A highly doped graphene sheet does support propagation of GP in the near-infrared window, while it is relatively difficult to excite the propagating GP [11, 12]. Due to easy control and manipulation of localized plasmons supported by graphene nanostructures, localized GPs have been widely explored in the mid-infrared and terahertz windows [13-15]. In graphene nanostructures (with D being the characteristic dimension), the plasmon-resonance wavelength associated with localized GPs scales as $\sqrt{E_F/D}$, where $E_F$ is the Fermi level of graphene [16]. Thus, shorter wavelength can be achieved through both higher doping levels and/or reduced structure dimensions. Until now, the shortest wavelength of the localized GP was observed around 3.7 μm in a 20 nm wide graphene nano-ring formed by using electron-beam lithography [17] this value is still quite far away from the wavelength window used in telecommunication. It has been suggested to use bottom-up approaches to control graphene features even with atomic-scale resolution [18, 19]. In this letter, we demonstrate the first realization of graphene nanostructures supporting localized GP near 2 μm by use of a block copolymer self-assembly method combined with oxygen reactive ion etching. We use a fully scalable block copolymer lithography process for preparation of highly uniform and periodic cylinder/sphere templates [20], which allow for a good control of size and uniformity of the produced graphene nanostructures.

Figure 1 shows the overall nanofabrication process. First the polystyrene-*b*-polydimethylsiloxanes (PS-PDMS) were directly spin-cast on a graphene/SiO$_2$/Si substrate without any preliminary surface modification (like e.g. the ubiquitous surface grafting of a brush layer [21] used in other studies). Selective (to the PS domain) solvent vapor annealing (SVA) was then applied to generate well-ordered vertical hexagonal cylinder or monolayer packed sphere morphology over a large area. The following O$_2$ dry etching enables simultaneous formation of the hard silicon oxycarbide nanocylinder/nanosphere by oxidation of the PDMS block, removal of the PS block and patterning of the underlying graphene. The size of graphene nanodisks can be tuned by carefully adjusting the dry etching time due to lateral etching effect.

The scanning-electron microscopy (SEM) and atomic-force microscopy (AFM) images of the nanocylinder array on different substrates are presented in Fig. 2. Silicon oxycarbide nanocylinders were formed from the Si-containing PDMS block under oxygen plasma [22]. A well-ordered hexagonally packed cylinder morphology is clearly observed in Figs. 2a-b.

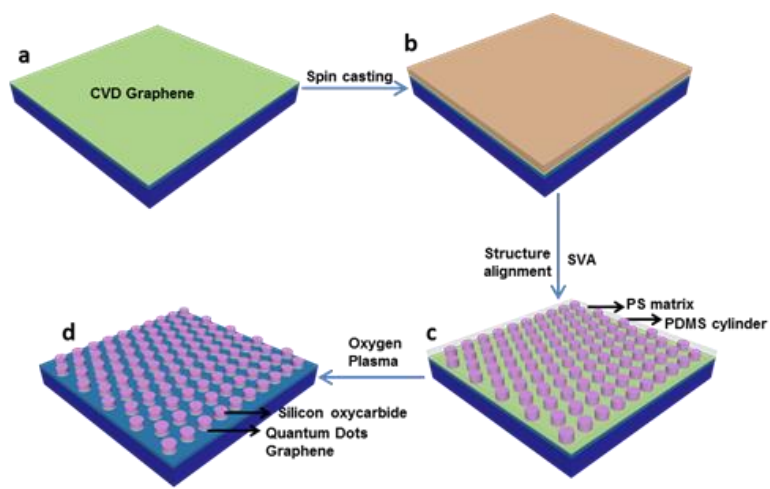

Fig. 1. Fabrication of graphene-nanodisk arrays (GNDAs) by direct block copolymer nanolithography. (a) CVD graphene on SiO2/Si substrate. (b) Spin coating of block copolymer thin film directly on CVD graphene. (c) Structure alignment of block copolymer via solvent vapor annealing. (d) Fabrication of hard silicon oxycarbide nanocylinder array through oxidation of PDMS, simultaneous removal of PS and etching of graphene under oxygen plasma.

Graphene sheets were patterned simultaneously during oxygen plasma etching on the Si/SiO$_2$ substrate. Compared to the silicon oxycarbide nanocylinders directly on Si/SiO$_2$ substrate, the order of nanocylinders on the graphene substrate is less well-defined (Fig. 2c, d) due to the graphene grain boundary and folds. The silicon oxycarbide mask can be removed by sonication in ethanol at room temperature, thus leaving an array of monolayer graphene nanodisks on the substrates (Fig. 2e, f). Note that graphene nanodisks are occasionally removed under the ultrasonication process; however, the remaining population of nanodisks is more than sufficient for a direct size characterization.

For the silicon oxycarbide mask, the diameter remains around 24 nm regardless of the etching time by oxygen plasma. In contrast, the diameter of graphene nanodisk can still be altered, e.g., from 25 to 18 nm due to lateral etching when the time for the oxygen plasma changes from 8 s to 10 s, see Fig. 3. We used an ImageJ plugin that extracts the dimensions from the SEM images (Fig. 3a, b) of disjoint nanodisk after identifying their shapes to evaluate their disks size distribution (Fig. 3c, d) [23]. Moreover, AFM images and profiles confirm that a GNDA monolayer is left on the substrate (Fig. 4a).

Here, we use Raman spectroscopy to examine the possible defects generated during the oxygen etching process. Fig. 4b shows the Raman spectra of GNDAs /Silicon oxycarbide nanodots after 0 s (black curve), 6 s (red curve), 8 s (blue curve) and 10 s (magenta curve) of oxygen plasma, revealing the evolution of defect formation. For the pristine CVD graphene used in this study, only a very small Raman D peak is observed, indicative of the good structural quality of graphene. After the etching in the presence of oxygen plasma, a prominent disorder-induced D peak appears at 1358 cm$^{-1}$. In addition, the double-resonance 2D peak becomes weaker. These observations suggest the presence of a larger number of defects related to the formation of nanodisk edges. Also the intensity ratio $I(D)/I(G)$ between Raman D and G peaks (commonly used to characterize disorder in graphene), is observed to increase with the time of oxygen plasma. When increasing etching time, We observe a significant increase of the $I(D)/I(G)$ ratio, from the initial average of 0.05 for 0 s to 0.72 for 6 s,

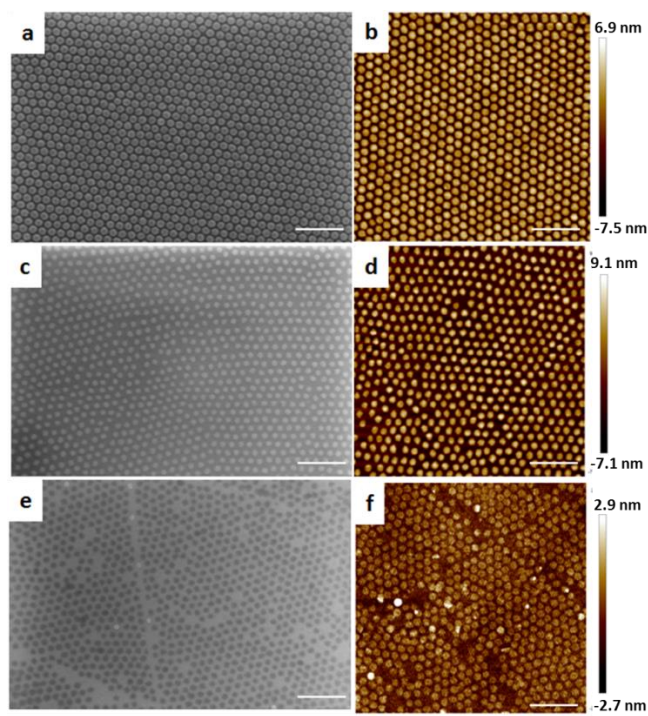

Fig. 2 SEM (left-column panels) and AFM (right-column panels) images of silicon oxycarbide nanocylinders mask on (a-b) SiO$_2$/Si substrate, (c-d) on graphene, and (e-f) GNDAs after removal of the mask by sonication. Scale bars: 200 nm.

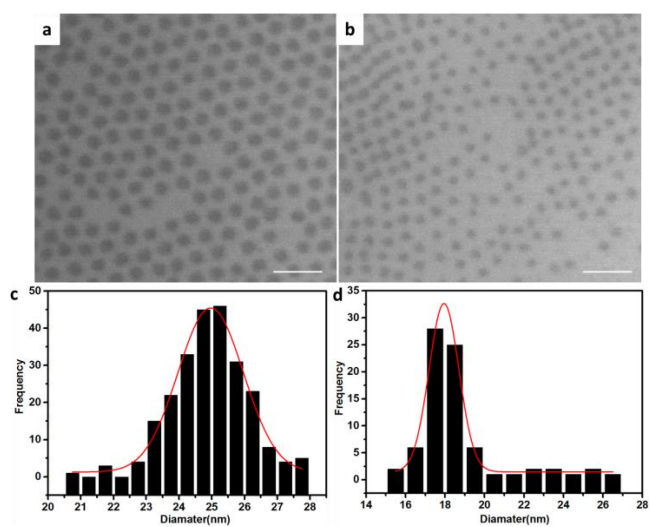

Fig. 3 SEM images of GNDAs after (a) 8 s and (b) 10 s of oxygen plasma etching. Size distribution of GNDAs after (c) 8 s and (d) 10 s of oxygen plasma etching. Scale bars: 100 nm.

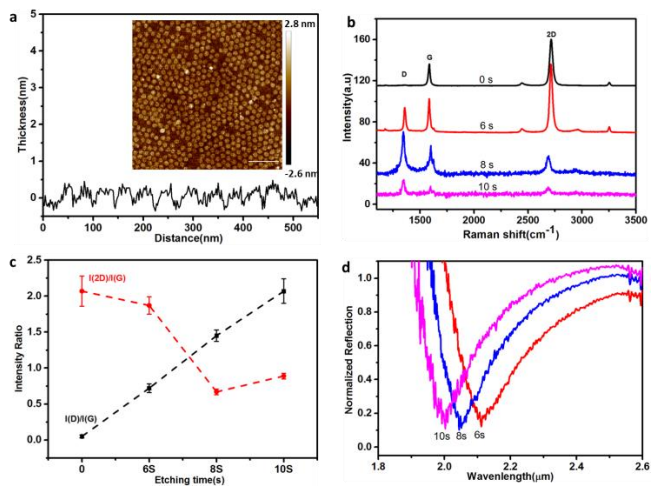

Fig. 4 (a) AFM line-scan profile of GNDAs on SiO$_2$/Si substrate after 8 s oxygen plasma. Inset: Corresponding AFM image. Scale bars: 200 nm. (b) Raman spectra, (c) Evolution of the peak intensity ratios I(D)/I(G) (black) and I(2D)/I(G) (red) and (d) normalized reflection spectra for the GNDAs/Silicon oxycarbide nanodisk formed after 6 s, 8 s and 10 s of oxygen plasma. As a reference, the Raman spectrum before oxygen etching (0 s) is shown by the black line in (b). The dips in (d) are associated with the excitation of localized GPs, leading to strong absorption.

1.45 for 8 s, and 1.55 for 10 s as shown in Fig. 4c. A higher defect density is also visible in the SEM image of the graphene etched for 10 s (Fig. 3b) as compared to the sample etched for 8 s (Fig. 3a).

We examine reflection spectra for the GNDAs using Fourier transform infrared spectroscopy (FTIR, Bruker VERTEX 70) at room temperature with the reflection angle of 20°. To exclude the influence of water vapor surrounding the sample, the system was purged with nitrogen overnight prior to measurement. The centimeter-scale size of the fabricated GNDAs allows us to measure with a relatively large light spot, giving a strong optical response. The unpolarized reflection spectra for the fabricated GNDAs are shown in Fig. 4d, where the unpolarized reflection for the bare graphene sheet on SiO$_2$/Si is taken as a reference. Prominent plasmonic resonances appear in the near-infrared region for these graphene nanodisks. The excitation of dipole resonances in the graphene nanodisk results in enhanced absorption, thus giving rise to significant dips in the reflection spectra. As the oxygen plasma etching time increases, the GP resonances are blue-shifted as shown in Fig. 4d. The blue-shift effect is consistent with the diminished size of GPs at longer etching time. We observe a clear 60 nm blue-shift, i.e. from 2.11 μm for 6 s oxygen plasma to 2.05 μm for 8 s oxygen plasma, and an additional 50 nm shift from 8 s to 10 s oxygen plasma.

The CVD-grown graphene is hole-doped, and the G mode frequency $\omega_G$ in the Raman spectra can be used to estimate the Fermi level in the graphene [24, 25] as $|E_F| = \frac{\omega_G - 1580\ cm^{-1}}{42\ cm^{-1}/eV}$. As indicated in Fig. 4b, the G mode frequency changes during the treatment by oxygen plasma, e.g. it shifts from 1583 cm-1 for the pristine graphene (0 s) to 1600 cm-1 after 10 s of plasma treatment, suggesting that the graphene is further hole-doped during the fabrication process. For the case of 10 s, $E_F$ is around -0.48 eV, while we estimate an initial Fermi level of 0.07 eV for 0 s. Note that when the size of graphene structures becomes smaller and smaller, we need to be concerned with atomic-scale details, such as quantum mechanical effects associated with electronic edge states [26-28]. Ignoring for simplicity such size/quantum effects, coupling effects for the neighboring nanodisks, and interband transitions, the GPP dipole-resonance supported by the graphene nanodisk can be evaluated as $\lambda = 2\pi c \sqrt{\frac{\pi \hbar^2 \varepsilon_0 (\varepsilon_1 + \varepsilon_2) D}{2e^2 E_F}}$, where c is the speed of light in vacuum, $\hbar$ is the reduced Planck constant, $\varepsilon_1$ and $\varepsilon_2$ are the permittivities of cladding and substrate, respectively. For our experiments, $\varepsilon_1$=1 and $\varepsilon_2$=2.1. Equipped with experimentally determined numbers for λ and $E_F$, we obtain D of ~4.6 nm (for the case of 10 s) by use of the classical approach. However, the estimated sizes are significantly smaller than the apparent physical dimensions (~18 nm) shown in Fig. 2d. We speculate that the longer etching time causes more defects within the nanodisks that cannot be observed by SEM or AFM. We note that in recent work on graphene plasmons in graphene ribbons [14], a significant reduction (~25%) of the apparent width of E-beam fabricated ribbons (with a subtraction of 28 nm) was applied in order to match well with the experimental results[14]. All these effects (including quantum effects, fabrication impurities and interband transition) can potentially explain the difference in terms of the size of the graphene nanodisks, but this requires further investigations beyond the mere measurement of optical spectra.

In summary, we have realized wafer-scale graphene nanodisk arrays with the aid of the block copolymer self-assembly method, where the physical diameter of the nanodisks are down to 18 nm. The dipole resonance of the graphene plasmon has been pushed down to 2.00 μm, which is to the best of our knowledge the smallest value reported for the localized GP resonance. The results obtained here can facilitate graphene plasmons both for fundamental study and for potential applications in the telecommunication window. The technique used here can even produce smaller structures by appropriate choice of the block copolymer composition and molecular weight.

**Funding.** This work was supported by the Danish National Research Foundation Center for Nanostructured Graphene, Project No. DNRF103.

**Acknowledgment**. We thank Dr. Bo-Hong Li for his technical assistant for FTIR measurement.